\documentclass[a4paper, USenglish]{article}
\usepackage{xspace}
\usepackage{color,xcolor}
\usepackage{graphicx}
\usepackage{tabularx}
\usepackage{algorithm,algorithmic}
\usepackage{amssymb}
\usepackage{amsmath}
\usepackage{booktabs} 
\usepackage{authblk}

\usepackage{tikz}
\usetikzlibrary{matrix,shapes,arrows,positioning,backgrounds,calc}
\usetikzlibrary{decorations.text}

\usepackage[hidelinks=true]{hyperref}
\usepackage[page]{appendix}

\newcommand{\etal}{\emph{et al.}\@\xspace}

\newcommand{\NONUMBER}[1]{\setcounter{ALC@rem}{1}\STATE #1{\addtocounter{ALC@line}{-1}\setcounter{ALC@rem}{0}}}
\newcommand{\FUNCTION}[1]{\STATE \textbf{procedure} #1 \begin{ALC@g}}
\newcommand{\ENDFUNCTION}{\end{ALC@g}\STATE \textbf{end procedure}}

\newcommand{\CASE}[1]{\STATE \textbf{case} #1\textbf{:} \begin{ALC@g}}
\newcommand{\ENDCASE}{\end{ALC@g}}

\newcommand{\DEFAULT}{\STATE \textbf{default:} \begin{ALC@g}}
\newcommand{\ENDDEFAULT}{\end{ALC@g}}
\newcommand{\DEFAULTLINE}[1]{\STATE \textbf{default:} }

\newtheorem{observation}{Observation}
\newtheorem{definition}{Definition}
\newtheorem{problem}{Problem}

\definecolor{aa}{HTML}{FFB338}
\definecolor{bb}{HTML}{FD8938}
\definecolor{cc}{HTML}{FC6139}
\definecolor{dd}{HTML}{FB3939}
\definecolor{ee}{HTML}{FA3A62}
\definecolor{ff}{HTML}{F93B89}
\definecolor{gg}{HTML}{F83BB0}
\definecolor{hh}{HTML}{F63CD6}
\definecolor{ii}{HTML}{F03DF5}
\definecolor{jj}{HTML}{C93DF4}
\definecolor{kk}{HTML}{B565F6}
\definecolor{ll}{HTML}{B490F7}
\definecolor{mm}{HTML}{9E8EF6}
\definecolor{nn}{HTML}{8D95F4}
\definecolor{oo}{HTML}{91ACF5}
\definecolor{pp}{HTML}{4192ED}
\definecolor{qq}{HTML}{41B5EC}
\definecolor{rr}{HTML}{42D7EB}
\definecolor{ss}{HTML}{42EADB}
\definecolor{tt}{HTML}{43E8B7}
\definecolor{uu}{HTML}{43E795}
\definecolor{vv}{HTML}{44E673}
\definecolor{ww}{HTML}{44E552}
\definecolor{xx}{HTML}{58E445}
\definecolor{yy}{HTML}{78E345}
\definecolor{zz}{HTML}{98E246}

\definecolor{00}{rgb}{0.7,0.7,0.7}
\definecolor{gr0}{rgb}{0.7,0.7,0.7}

\newcommand{\telstart}[2]{
    \pgfmathsetmacro{\x}{#1}
    \pgfmathsetmacro{\y}{#2}
    \fill[color=gray] (\x+0.18, \y+0.16) arc (90:270:0.16 and 0.16) -- (\x+0.18,
    \y+0.16);
}
\newcommand{\telend}[2]{
    \pgfmathsetmacro{\x}{#1}
    \pgfmathsetmacro{\y}{#2}
    \fill[color=gray] (\x, \y+0.16) arc (270:90:-0.16 and -0.16) -- (\x,
    \y+0.16);
}

\newcommand{\charupb}[3]{
    \pgfmathsetmacro{\x}{#1}
    \pgfmathsetmacro{\y}{#2}
    \foreach \char [count=\ci] in #3 {
        \fill[draw=none,color=\char\char] (\x, \y +\ci*0.32 - 0.16) rectangle
        (\x+0.78, \y +\ci*0.32 - 0.48);
        \node[color=black] at (\x+0.39, \y + \ci*0.32 - 0.32) {\textbf{\small\char}};
    }
}

\title{Fast computation of approximate weak common intervals in multiple
indeterminate strings}

\author[1]{Daniel Doerr}
\author[2]{Bernard M.E. Moret}

\affil[1]{Institute for Medical Biometry and Bioinformatics, Heinrich Heine
University, D\"usseldorf, Germany}

\affil[2]{School of Computer and Communication Sciences, EPFL, CH-1015,
Lausanne, Switzerland}

\date{}

\begin{document}

\maketitle

\begin{abstract}
   In ongoing work to define a principled method for syntenic block discovery
   and structuring, work based on homology-derived constraints and a
   generalization of common intervals, we faced a fundamental computational
   problem: how to determine quickly, among a set of indeterminate strings
   (strings whose elements consist of subsets of characters), contiguous
   intervals that would share a vast majority of their elements, but allow for
   sharing subsets of characters subsumed by others, and also for certain
   elements to be missing from certain genomes. An algorithm for this problem in
   the special case of determinate strings (where each element is a single
   character of the alphabet, i.e., ``normal" strings) was described by Doerr
   et al., but its running time would explode if generalized to indeterminate
   strings. In this paper, we describe an algorithm for computing these special
   common intervals in time close to that of the simpler algorithm of Doerr et al.\ 
   and show that can compute these intervals in just a couple of hours for
   large collections (tens to hundreds) of bacterial genomes.
\end{abstract}

\section{Introduction}
The rapidly increasing number of whole-genome sequences in public repositories
is creating a demand for an effective means of comparing multiple whole genomes.
Such comparisons not only enrich our knowledge about present-day organisms, they
also open a window into their evolutionary past.   Many genomes of interest
(such as vertebrate genomes) are much too large for base-pair by base-pair
comparisons and such comparisons are made all the harder when multiple genomes
are being considered.   Thus large-scale whole-genome comparative studies
are limited by two factors: (i) the size of the genomes (typically in the billions
of base pairs for vertebrates) and (ii) the number of genomes one wishes to compare.
The standard approach to the first problem over the last 15 years has been to
decompose the genomes into \emph{syntenic blocks}.  In their most common form,
syntenic blocks are contiguous blocks ranging from $10^4$ to $10^7$ base pairs
that form clearly conserved units across the genomes---as attested by the conservation
of closely related sets of \emph{genomic markers}.   Rather than comparing genomes
base-pair by base-pair, or even marker by marker, one compares them syntenic
block by syntenic block.   In other words, genomes with billions of base pairs
are represented by several thousands of syntenic blocks.  As most of the work
on such decompositions was carried out using genes as genomic markers, one
can view the ordering of syntenic blocks along each genome as a generalization
of the ordering of genes along each genome: just as genes in each genome are
homologous to genes in other genomes and form gene families, so do syntenic blocks.
(The other problem, of having to compare multiple genomes, is perhaps
best carried out in a phylogenetic context, where direct comparisons, the number of which
grows quadratically with the number of genomes, are eschewed in favor of comparisons
between descendants and common ancestors \cite{ZhangMoret}, but it is not
our focus here.)

In ongoing work to define a principled method for syntenic block discovery
and structuring, work based on homology-derived constraints \cite{Ghiurcuta:2014dx}
and a generalization of \emph{common intervals} \cite{Uno:2000ca}, we faced a
fundamental computational problem: how to determine quickly, among a set
of indeterminate strings (strings whose elements consist of subsets of characters),
contiguous intervals that would share a vast majority of their elements, but allow
for sharing subsets of characters subsumed by others, and also for certain elements
to be missing for certain genomes.  An algorithm for this problem in the special case
of determinate strings (where each element is a single character of the alphabet,
i.e., ``normal" strings) was described by Doerr \emph{et al.}\ \cite{Doerr:2014fq},
but its running time would explode if generalized in the ``obvious" way to
indeterminate strings.  In this paper, we describe an algorithm for computing
these special common intervals in time close to that of the simpler algorithm of
Doerr \emph{et al.}\ \cite{Doerr:2014fq} that makes it possible to compute
these intervals at the scale of complete vertebrate genomes.

\section{Background}
We begin with a quick sketch of our approach to the discovery of syntenic blocks
in  multiple genomes, to put in proper perspective the computational problem that is 
the focus of this paper.

The identification of large conserved regions across genomes, i.e., syntenic blocks,
is in itself useful, as it allows us to circumscribe regions of interest and focus
our studies on these.  However, syntenic blocks can be also be used to describe
whole genomes at a higher level of abstraction, enabling further analyses to be
conducted at this new level.  Indeed, the number of levels of decomposition
(or abstraction) need not be limited to two or three: at one extreme, each position
in a genomic sequence can be viewed as a syntenic block while, at the other extreme,
each whole genome can be viewed as a single syntenic block.    Neither extreme
is of much interest and the study of short subsequences is well advanced, so
our design starts at the level of \emph{genomic markers}, that is, short sequences
that are both short enough to be found in (exactly or much) the same form in other
genomes and long enough that their conservation is statistically significant.
Such markers can be obtained through genome segmentation~\cite{Visnovska:2013ua}
or, more commonly, using already available annotations of the genomic sequences
that mark conserved regions such as noncoding and protein-coding genes.

Our method relies on predefined \emph{homologies} between these genomic markers.
(Recall that homology is an equivalence relation that denotes shared ancestry.)\ \ 
Based on homologies between genomic markers, we derive homologies between syntenic
blocks using the formal criteria first enunciated by Ghiurcuta and
Moret~\cite{Ghiurcuta:2014dx}.  Informally, according to their
criterion, two syntenic blocks are homologous if each of their constituent
genomic markers is homologous to a genomic marker in the opposite block. 

From a conceptual point of view, shared origin is a common feature of
genomic markers; in fact, a perfect oracle for homology would not provide
very useful information---it would lack differentiability and thus be
ineffective for synteny analysis.
It is thus fortunate, in a way, that the divergence of sequences through
mutational changes renders inference of homology impossible after a certain
amount of time.   Unfortunately, this same divergence renders the process of
homology inference error-prone and unlikely to result in an equivalence
relation.   Most current algorithms for homology inference rely on some
type of clustering, often based almost entirely on sequence similarity,
but homology should also take into account relative placements within each
genome.   In other words, homology assignments between markers are required
for synteny analysis, yet their embedding in syntenic blocks provides
a powerful indicator of true homology.   We break this circularity
by not requiring that the inferred homologies describe an equivalence
relation: we drop the requirement of transitivity.  This relaxation leads
to a decisive advantage in practice: genomic markers no longer need to be
clustered into families, so that the synteny analysis can be directly
performed on pairwise sequence similarities. 

We address the problem of identifying syntenic blocks under weak homology
by identifying sets of \emph{approximate weak common intervals}
(AWCIs) in indeterminate strings---and this is the computational problem
we address in this paper.  Informally, indeterminate strings are
sequences in which each position is associated with a (nonempty) subset
of alphabet characters.  Two intervals in two indeterminate strings
are \emph{weak common intervals} if each position of both intervals
contains at least one character that is also present in the other interval.
If a character encoding for two indeterminate strings represents (non-transitive)
homologies between markers of two genome sequences, weak common intervals are
equivalent to homologous syntenic blocks. Using AWCIs, we identify syntenic
blocks that mildly violate the synteny criterion by allowing for a limited
number of missing homology statements. We do this to account for (i) false
negative errors in the inferred homology assignment and (ii) insertions or
deletions of one or a few genomic markers that are a result of genome evolution. 

\section{A First Algorithm}

\subsection{Basic definitions}
We recall some basic definitions of common intervals in indeterminate strings from
Doerr~\etal~\cite{Doerr:2014fq}.
An indeterminate string $S$ with $n$ index positions is a string
over the power set $\mathcal P(\Sigma) \setminus \emptyset$ of an alphabet
$\Sigma$. That is, for each $i$, $1\leq i\leq n$, we have $S[i]\subseteq\Sigma$ and
$S[i]\neq\emptyset$, where $S[i]$ denotes the character set associated with
the $i$-th position in $S$.  We denote the \emph{length} of an indeterminate string $S$
with $n$ index positions by $|S| \equiv n$ and its \emph{cardinality}, i.e.,
the number of \emph{all} elements in $S$, by $\|S\| \equiv \sum_{i=1}^{n} |S[i]|$. Two
positions $i$ and $j$, $1 \leq i \leq j \leq |S|$, induce the (indeterminate)
\emph{substring} $S[i, j] \equiv S[i]\,S[i+1]\, \dots\,S[j]$. To distinguish
intervals in different indeterminate strings, we indicate the affiliation of an
interval $[i,j]$ to indeterminate string $S$ by the subscript notation
$[i,j]_S$.

Common intervals were first defined on permutations~\cite{Uno:2000ca} and subsequently
extended to ordinary, then indeterminate, strings~\cite{Didier:2007hj,Doerr:2014fq}.
The idea behind common intervals is to compare substrings based on their character sets.
The \emph{character set} of an indeterminate string $S$ is defined as
$\mathcal C(S) \equiv \bigcup_{i=1}^{n} S[i]$. 
Given two indeterminate strings $S$ and $T$, two intervals, $[i, j]$ in $S$ and
$[k, l]$ in $T$, are \emph{weak common intervals} with \emph{common character
set} $C=\mathcal C(S[i,j]) \cap \mathcal C(T[k,l])$ if for each $i'$, $i\leq
i'\leq j$, we have $C \cap S[i'] \neq \emptyset$ and for each $k'$, $k \leq
k'\leq l$, we have $C \cap T[k'] \neq \emptyset$.  We study a
variant of weak common intervals that tolerates a limited number of insertions
and deletions.  Given some threshold $\delta \geq 0$, two
intervals, $[i,j]$ in indeterminate string $S$ and $[k,l]$ in indeterminate
string $T$, are ($\delta$-)\emph{approximate weak common intervals} with
\emph{common character set} $C=\mathcal C(S[i,j]) \cap \mathcal C(T[k,l])$ if
the number of \emph{positions} with no intersection with $C$ is limited by
$\delta$, i.e., if we can write
  $$|\{x\mid i\leq x\leq j\colon S[x]\cap C = \emptyset\}|+|\{y~|~k\leq y\leq l\colon T[y]\cap C = \emptyset\}|\leq\delta$$
We shall call these positions \emph{indels}.
A set of intervals is a set of \emph{approximate weak common intervals} if every
pair of members satisfies the property.

\begin{figure}[tb]%
\begin{center}%
    \begin{tikzpicture}[]%
        \draw[draw=none,fill,color=black,rounded corners=3](-10,10.16)
        rectangle (-3.62, 9.7);
        \draw[draw=none,fill,color=black,rounded corners=3](-9.2,8.96) rectangle
        (-4.42, 8.5);
        \draw[draw=none,fill,color=black,rounded corners=3](-10,7.76) rectangle
        (-3.62, 7.3);
%

        \node[anchor=east] at (-10.2, 10) {$\mathbf{S_1}$};
        \telstart{-10.2}{10};
        \charupb{-10}{10}{{g}};
        \charupb{-9.2}{10}{{b,p}};
        \charupb{-8.4}{10}{{x}};
        \charupb{-7.6}{10}{{n,p}};
        \charupb{-6.8}{10}{{d,o,s}};
        \charupb{-6}{10}{{a,z}};
        \charupb{-5.2}{10}{{e,w}};
        \charupb{-4.4}{10}{{f}};
        \charupb{-3.6}{10}{{v,l}};
        \charupb{-2.8}{10}{{h,u,z}};
        \charupb{-2}{10}{{j,r}};
        \charupb{-1.2}{10}{{k}};
        \telend{-0.4}{10};

        \node[anchor=east] at (-10.2, 8.8) {$\mathbf{S_2}$};
        \telstart{-10.2}{8.8};
        \charupb{-10}{8.8}{{c,k}};
        \charupb{-9.2}{8.8}{{f,n,p}};
        \charupb{-8.4}{8.8}{{w}};
        \charupb{-7.6}{8.8}{{b,d}};
        \charupb{-6.8}{8.8}{{x}};
        \charupb{-6}{8.8}{{c,l,m}};
        \charupb{-5.2}{8.8}{{a,g}};
        \charupb{-4.4}{8.8}{{r}};
        \charupb{-3.6}{8.8}{{a,w,x}};
        \charupb{-2.8}{8.8}{{p}};
        \charupb{-2}{8.8}{{f,z}};
        \telend{-1.2}{8.8};

        \node[anchor=east] at (-10.2, 7.6) {$\mathbf{S_3}$};
        \telstart{-10.2}{7.6};
        \charupb{-10}{7.6}{{d}};
        \charupb{-9.2}{7.6}{{g,b}};
        \charupb{-8.4}{7.6}{{a}};
        \charupb{-7.6}{7.6}{{p,s}};
        \charupb{-6.8}{7.6}{{n}};
        \charupb{-6}{7.6}{{a,b}};
        \charupb{-5.2}{7.6}{{f,m,w}};
        \charupb{-4.4}{7.6}{{e,w}};
        \charupb{-3.6}{7.6}{{k}};
        \charupb{-2.8}{7.6}{{j,u}};
        \charupb{-2}{7.6}{{h}};
        \charupb{-1.2}{7.6}{{c,r}};
        \charupb{-0.4}{7.6}{{z}};
        \telend{0.4}{7.6};
    \end{tikzpicture}%
  \end{center}%
  \caption{Example of three indeterminate strings $S_1$, $S_2$, $S_3$. The black
     underlined intervals form a closed set of $1$-approximate weak common intervals.}%
  \label{fig:wci}%
\end{figure}
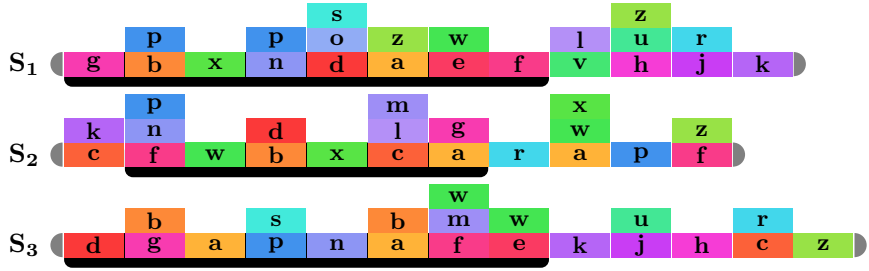

Finally, we extend the property of pairs of \emph{mutually closed} intervals
of~\cite{Doerr:2014fq} to sets of two or more intervals: 
\begin{definition}[closed set of intervals]\label{def:closed}
    A set $\mathcal I$ of intervals is \emph{closed} if each interval
    $[i,j]_S\in\mathcal I$ has neither immediate left nor immediate
    right neighboring position $p$ with $S[p]\cap \mathcal C(S'[k, l])$
    for each $[k, l]_{S'}\in\mathcal S\setminus\{[i,j]_S\}$. 
\end{definition}
Note that closedness is a non-hereditary property of sets of intervals: a subset
of a closed set of intervals need not be closed itself. Figure~\ref{fig:wci}
shows an example of a closed set of $1$-approximate weak common intervals. 

\section{Identifying Maximal Closed Sets of AWCIs}\label{sec:speedup} 
A set is \emph{maximal} if it cannot be extended by including
further valid members.
\begin{problem}[maximal closed sets]%
    \label{prb:awci}%
    Given $m$ indeterminate strings $S_1,\ldots,S_m$ and indel threshold
    $\delta\geq 0$, discover all maximal closed sets of $\delta$-approximate
    weak common intervals that have members in at least $q$ out of the $m$
    indeterminate strings.
\end{problem}%
Because closedness of sets of intervals does not extend to subsets,
each set of AWCIs that is identified in the course of the enumeration
needs to be tested individually for this property.  For the same reason,
the closed pairs of AWCIs returned by the discovery algorithms described
in~\cite{Doerr:2014fq} cannot be used for the construction of larger sets
of AWCIs in multiple sequences, because non-closed pairs are omitted
that could contribute to larger closed sets of AWCIs.

Since there can be exponentially many sets of AWCIs in a given set of
indeterminate strings, we do not enumerate them explicitly, but instead
construct a compact representation from which they can be extracted.  Given a
set of indeterminate strings, $\{S_1,\ldots,S_m\}$, we construct a graph
$G(S_1,\ldots,S_m)$, where each vertex corresponds to an interval in one of the
indeterminate strings and any two vertices are connected by an undirected edge
if and only if their corresponding intervals form a pair of AWCIs.  Identifying
maximal closed sets of AWCIs then reduces to finding within the graph maximal
cliques of vertices whose corresponding AWCIs are closed. In constructing the
graph, we already omit intervals that cannot participate in sets of AWCIs of
minimum size $q$.  We test each vertex for possible participation in a closed
set of AWCIs.  That is, if the neighbors of a vertex $v$ are all connected to a
vertex $u$ such that the interval $I_v$ associated to vertex $v$ is a proper
subinterval of the interval $I_u$ corresponding to vertex $u$, and if interval
$I_u$ extends to at most one position left and/or right of $I_v$ that shares
characters with neighbors of $I_v$, then vertex $v$ can be discarded.  The
resulting graph is enriched with maximal closed cliques of AWCIs.

\subsection{MACSI -- An extension of ACSI}\label{sec:macsi} 
We now describe MACSI, a simple approach for enumerating AWCIs in indeterminate
strings---see Algorithm~\ref{alg:macsi}.
\begin{algorithm}[b]%
    \caption{MACSI enumerates all paris of approximate weak common intervals
    that are shared by at least $q$ out of $m$ indeterminate strings.}%
    \label{alg:macsi}%
{\footnotesize
\begin{algorithmic}[1]%
    \REQUIRE Indeterminate strings $S_1, \ldots, S_m$, indel threshold
    $\delta$, quorum $q$.
    \ENSURE All approximate weak common intervals of $S_1, \ldots, S_m$ with at
    most $\delta$ indels that are conserved in at least $q$ strings
    \vspace{0.5em}
    \STATE Construct character set intersection tables $\textsc{Pos}_{xy}$ for
    all sequence pairs $\{x, y\} \subseteq [1, m]$ 
    \FOR{$x\gets 1~\TO~m-1$} \label{alg:l_ref_start}
        \FOR{$i \gets 1~\TO~|S_x|$} \label{alg:l_left_bound_start}
            \STATE Determine set $J$ of potential right bounds of AWCI
            pairs with left bound $i$ in $S_x$ \label{alg:l_det_j}
            \FOR {\textbf{each } $j$ \textbf{in} $J$}
                \STATE Initialize empty list \textsc{Ints}
                \FOR{$y \gets x+1~\TO~m$} \label{alg:l_y_start}
                    \STATE construct sorted set $P_y$ of positions
                    $\textsc{Pos}_{xy}[i] \cup \cdots \cup
                    \textsc{Pos}_{xy}[i+\delta]$ \label{alg:l_con_p}
                    \STATE $p_{\mathit{prev}} \gets -1$
                    \FOR{\textbf{each } $p$ \textbf{in} $P_y$} \label{alg:l_iter_p_start}
                        \STATE Append to list \textsc{Ints} all intervals
                        $[k, l]$ in $S_y$, $p_{\mathit{prev}} < k \leq p
                        \leq l$, that are approximate weak common intervals
                        with $[i, j]$ in $S_x$ \label{alg:l_int_kl}
                        \STATE $p_{\mathit{prev}} \gets p$
                    \ENDFOR \label{alg:l_iter_p_end}
                \ENDFOR \label{alg:l_y_end}
                \IF {\textsc{Ints} contains intervals from at least $q-y+1$
                sequences}
                    \STATE \textbf{report} $\{([i, j]_{S_x}, [k,
                    l]_{S_y})~|~[k, l]_{S_y} \in \textsc{Ints} \}$
                \ENDIF
            \ENDFOR
        \ENDFOR \label{alg:l_ref_end}
    \ENDFOR \label{alg:l_left_bound_end}
    \vspace{1em}%
\end{algorithmic}%
}%
\end{algorithm}%
MACSI is a simple extension of Algorithm ACSI described by
Doerr~\etal~\cite{Doerr:2014fq}; however, unlike ACSI, it enumerates
any pairs of AWCIs rather than only closed ones. 

Instead of repeatedly determining intersections of character sets across
positions of distinct indeterminate strings, the positions of intersecting
character sets are identified and stored at the beginning.  Given two
indeterminate strings $S_x$ and $S_y$, each position $j$, $1\leq j\leq |S_y|$ in
intersection table $\textsc{Pos}_{xy}$ is associated with a sorted list of
positions in $S_y$ that have a nonempty intersection with character
set~$S_x[j]$.

Given a set of sequences $S_1, \ldots, S_m$, each of its members acts once as
\emph{reference} (lines \ref{alg:l_ref_start}-\ref{alg:l_ref_end}), over which
the algorithm iterates, thereby fixing a left bound $i=1,\ldots,|S_x|$ (lines
\ref{alg:l_left_bound_start}-\ref{alg:l_left_bound_end}). The algorithm then
determines candidates for right bounds $J\subseteq\{i+1,\ldots,|S_x|\}$, where
each candidate interval $[i,j]_{S_x}$, $j\in J$, is potentially part of a set of
AWCIs conserved in at least $q$ out of the $m$ sequences.  Once the interval
$[i,j]$ in reference string $S_x$ is fixed, the algorithm iterates over the
remaining $m-x$ sequences to enumerate all such candidate intervals~(lines
\ref{alg:l_y_start}-\ref{alg:l_y_end}). 

The construction of set $J$ representing potential right interval bounds
in $S_x$ (line~\ref{alg:l_det_j}) is easily done in time linear in $|S_x|$.
For each fixed interval $[i, j]$ in $S_x$, the enumeration of all intervals in
$S_y$ that are AWCIs with $[i,j]_{S_x}$ can be achieved in $\mathcal O(|S_y|\cdot\|S_y\|)$
amortized running time.  The algorithm first collects and sorts all positions
$P_y\in S_y$ around which such intervals may lie (line \ref{alg:l_con_p}).
For each such position $p\in P_y$, all intervals with left bound larger
than the previously processed position are identified (lines
\ref{alg:l_iter_p_start}-\ref{alg:l_iter_p_end}). Testing for the approximate
common interval property of an interval pair $[i,j]_{S_x},[k,l]_{S_y}$, $k\leq p\leq l$,
can be done in $\mathcal O(1)$ by keeping track of observed indices of $[i,j]_{S_x}$, 
  $$C_{kl} = \{i' \in \textsc{Pos}_{yx}[k]\cup\textsc{Pos}_{yx}[l]~|~i\leq i'\leq j\}$$
and of the number of positions $d_{kl}$ in $[k,l]_{S_y}$ whose character sets do not
intersect with $[i,j]_{S_x}$.  Then $[i,j]_{S_x},[k,l]_{S_y}$ are AWCIs iff
$j-i+1-|C_{kl}|+d_{kl} \leq \delta$ holds and the enumeration of all intervals that
are AWCIs with $[i,j]_{S_x}$ can be done in $\mathcal O(|S_y|\cdot\|S_y\|)$ time. 

The overall running time of Algorithm~\ref{alg:macsi} is in $\mathcal O(m^2\cdot
n^2\cdot N^2)$, with $n=\max_{x=1..m} |S_x|$ and $N=\max_{x=1..m} \|S_x\|$. This
is close to optimal, since the number of pairwise AWCIs can be as large as the
total number of interval pairs, itself in $\mathcal O(m^2\cdot n^4)$. Hereby we
make the reasonable assumption that the cardinality $\|S\|$ of an inteterminate
string $S$ is in the order of $a\cdot|S|$ with $a \ll \Sigma$. 

\section{Speeding Up the Enumeration of AWCIs for Synteny Analysis}
We now study in more detail the construction of set $J$ (line~\ref{alg:l_det_j}),
which contains potential right interval bounds in $S_x$, and describe an
approach that significantly improves the running time by taking advantage
of two characteristics of genomes.   First, there exist many characters
shared only between a subset of all sequences; and second, each sequence
in the dataset is associated with a single genome representing a
concatenation of one or more chromosomes or contigs, yet AWCIs are prohibited
from spanning more than one.
As a result, our sequences exhibit a non-negligible number of indels that can
be trivially identified and used to preempt or terminate unsuccessful searches
of sets of AWCIs.

Recall that we are interested in identifying sets of AWCIs with members in a quorum of
at least $q \leq m$ strings.  Sets of AWCIs enforce transitivity,
but the AWCI property itself is not transitive. Therefore, even if we
identify a set of intervals $I_{x+1..m}$ with members in at least $q-1$ strings
and which form AWCI pairs with a given interval $[i,j]_{S_x}$, we cannot
guarantee that the intervals $I_{x+1..m}$ form AWCI pairs among themselves.
Thus, for each interval $[i,j]_{S_x}$, we want to assess whether it could
participate in at least one set of AWCIs that satisfies the quorum parameter
prior to constructing set $I_{x+1..m}$.
\begin{observation}
    Given sequences $S_1,\ldots,S_m$, an indel threshold $\delta$, and a quorum~$q$,
    an interval $[i,j]$ in $S_x$, for $1\leq x\leq m$, is \emph{not} a candidate
    for an AWCI pair if we have
    $|\{ |\mathcal C(S_x[i,j])\setminus\mathcal C(S_y)|\leq\delta~|~y\in [1,m]\setminus x\}|<q-1$.
\end{observation}
To exploit this observation, we construct two types of tables,
$\textsc{Ridge}_{xy}^c$ and $\textsc{Ridge}_{xy}^t$, for $\{x, y\}\subseteq [1,m]$,
where superscripts $c$ and $t$ stand for \emph{cis} and \emph{trans}, respectively.
The first, $\textsc{Ridge}^c_{xy}$, holds information within intersecton table
$\textsc{Pos}_{xy}$ to decide whether $[i, j]_{S_x}$ putatively participates in
an AWCI pair, whereas $\textsc{Ridge}^t_{xy}$ holds information from
$\textsc{Ridge}_{yx}^c$, allowing to test whether interval $[i, j]_{S_x}$ is
associated to intervals in $S_y$, with whom it potentially forms AWCI pairs.
Each position $0\leq p<|S_x|$ in table $\textsc{Ridge}_{xy}^c$ represents a
count of the number of positions $0\leq p'\leq p$ for which $S_x[p']$
contains no characters of $\mathcal C(S_y)$.
Thus interval $[i,j]$ in string $S_x$ cannot participate in any AWCI pair if 
$\textsc{Ridge}_{xy}^c[j]-\textsc{Ridge}_{xy}^c[i-1] > \delta$.

\begin{figure}[tb]%
    \centering \includegraphics[width=\columnwidth]{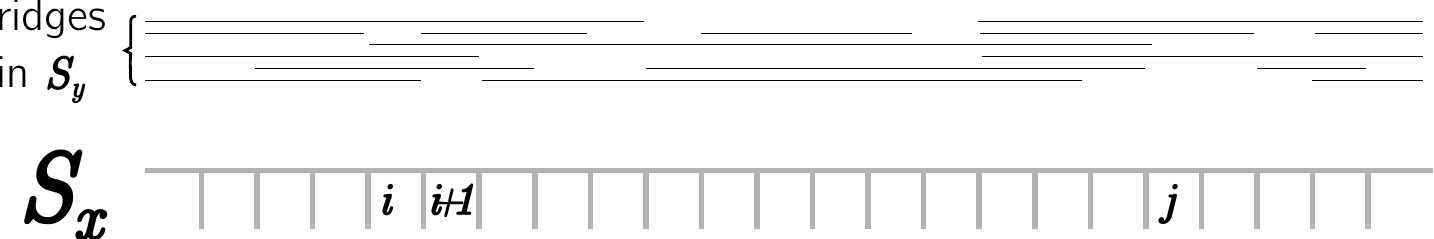}%
    \caption{The data structure $\textsc{Ridge}_{xy}^t$;
       lines indicate ridges, i.e., 1s across binary vectors
       $\textsc{Ridge}_{xy}^t[i']$ for some $i' \in [0,|S_x|]$.}%
    \label{fig:ridges}%
\end{figure}

We now construct a similar test for trans pairs of our reference interval
$[i,j]_{S_x}$ with the help of tables $\textsc{Ridge}_{xy}^t$.
We say that two positions $i,j$ of string $S_x$, for $0\leq i<j\leq|S_x|$,
reside on the same \emph{ridge} iff
$\textsc{Ridge}_{xy}^c[j]-\textsc{Ridge}_{xy}^c[i-1] \leq \delta$ holds. 
For a trans string
$S_y$, $x\neq y$, observe that the characters of $\mathcal C(S_x[i,j])$ can
occur in many different positions in $S_y$ and thus can be associated
with different intervals that form an AWCI pair with $[i,j]_{S_x}$. For each
candidate interval $[k,l]$ in $S_y$ we must have
$\textsc{Ridge}_{yx}^c[k]-\textsc{Ridge}_{yx}^c[l-1]\leq\delta$.
To test this property quickly, we construct binary vectors to identify
ridges in $S_y$ containing characters of $\mathcal C(S_x[i,j])$; an example
is shown in Figure~\ref{fig:ridges}.
Each index in any vector associated with positions $i..j$ in $S_x$
corresponds to a distinct ridge in $S_y$.  In order to test whether any
two positions $i'$, $j'$, $i\leq i'< j'\leq j$, contain characters that are
located on common ridges in $S_y$, it then suffices to perform a simple bitwise AND
operation between its associated binary vectors. Moreover, using a bitwise AND
of all vectors associated with positions in $i..j$ in $S_y$ we can decide
whether there exists at least one ridge in $S_y$ that contains characters
from all character sets $S_x[i']$, $i\leq i'\leq j$.

Data structure $\textsc{Ridge}_{xy}^t$ is of size $\mathcal O( (\delta+1) \cdot
|S_x|\cdot\|S_y\|)$ and its use increases the overall running time of MACSI by a
factor of $(\delta+1)$. In practice, its size is far from its asymptotic bound
and MACSI gains a significant speed-up. To minimize the size of these binary
vectors in $\textsc{Ridge}_{xy}^t$, we relax our requirement that each ridge be
assigned a unique index across all binary vectors.  We can do this without
damaging the consistency of our data structure, since we do not compare
arbitrary binary vectors against each other, but only those whose positions in
$S_x$ are at most $\delta$ trivial indels afar.  Therefore we reuse indices of
ridges in the binary vectors. Specifically, when constructing the data structure
$\textsc{Ridge}_{xy}^t$ from left to right, at a given position $j$ we assign a
newly observed ridge $\textsc{Ridge}_{yx}^c[k]$ (for some $k \in
\textsc{Pos}_{xy}[j]$) to a previously used index in the binary vector if the
ridge associated to this index has not been observed in any of the previous
positions in $S_x$ that are no more than $\delta$ indels afar. 

In Algorithm~\ref{alg:macsi_filter},
\begin{algorithm}[b]%
    \caption{Testing a position $j$ in string $S_x$ for the existence
    of some interval $[i',j']_{S_x}$, $i'\leq i\leq j\leq j'$---, i.e.,
    AWCIs with intervals in at least $q-1$ other indeterminate strings.}%
    \label{alg:macsi_filter}%
{\footnotesize
\begin{algorithmic}[1]%
    \REQUIRE start $i$ and current position $j$ in string $S_x$, tables
    $\textsc{Ridge}_{xy}^c$ and $\textsc{Ridge}_{xy}^t$, temporary data
    structures $\textsc{Active}^t$ and $\Delta^t$, indel treshold
    $\delta$, quorum $q$, reference $x$.
    \ENSURE Returns \textbf{True} if $S_x[i,j]$ is a candidate for
    an AWCI set in at least $q$ strings with less than $\delta$ indels and
    \textbf{False} otherwise.
    \vspace{0.5em}
    \STATE $\mathit{candidates} \gets 0$
    \FOR{$y \textbf{ in } [1,k] \setminus x$}
        \NONUMBER \COMMENT{\textit{Determine the location of currently masked
        (trans-) ridges in $\Delta^t[y]$}}
        \STATE $d \gets \max(0,\delta + 1 + \textsc{Ridge}_{xy}^c[i] -
        \textsc{Ridge}_{xy}^c[j])$ \label{alg:l_d_indels}
        \NONUMBER \COMMENT{\textit{Ridges that have been observed before
        but are not present at this position}}
        \STATE $a_{\mathit{off}} \gets \textsc{Active}^t[y] \land  \lnot
        (\textsc{Ridge}_{xy}^t[j] \lor \Delta^t[y][d])$ \label{alg:l_aoff}
        \NONUMBER \COMMENT{\textit{Ridges that have not been observed before but are present in
        $\textsc{Ridge}_{xy}^t[j]$}}
        \STATE $a_{\mathit{new}} \gets \textsc{Ridge}_{xy}^t[j] \land \lnot
        (\textsc{Active}^t[y] \lor \Delta^t[y][d])$ \label{alg:l_anew}
        \NONUMBER \COMMENT{\textit{Update $\Delta^t[y]$ for all ridges in
        $a_{\mathit{off}}$ and $a_{\mathit{new}}$}}
        \FOR {$d' \gets 0~\TO~d$} \label{alg:l_delta_update_start}
            \STATE $tmp \gets \Delta^t[y][d'] \land a_{\mathit{off}}$
            \STATE $\Delta^t[y][d'] \gets \Delta^t[y][d'] \lor a_{\mathit{off}}$
            \NONUMBER \COMMENT{\textit{Carry ridges of
            $a_{\mathit{off}}$ already marked in $\Delta^t[y][d']$ over to
            $\Delta^t[y][d'+1]$}}
            \STATE $a_{\mathit{off}} \gets tmp$
            \IF {$d' < j-i$}
                \STATE $\Delta^t[y][d'] \gets \Delta^t[y][d'] \lor a_{\mathit{new}}$
            \ENDIF
        \ENDFOR \label{alg:l_delta_update_end}
        \NONUMBER \COMMENT{\textit{Add ridges of current position to
        $\textsc{Active}^t[y]$}}
        \STATE $\textsc{Active}^t[y] \gets \textsc{Active}^t[y] \lor
        \textsc{Ridge}_{xy}^t[j]$ \label{alg:l_active_update}
        \NONUMBER \COMMENT{\textit{Test if $S_x[i,j]$ has at least one ridge
        $S_y$ with less than $\delta$  indels}}
        \IF{$\textsc{Active}^t[y] \land \lnot \Delta^t[y][d] \neq 0$}
        \label{alg:l_indel_test_start}
            \STATE $\mathit{candidates} \gets \mathit{candidates} + 1$
        \ENDIF \label{alg:l_indel_test_end}
    \ENDFOR
    \STATE \textbf{return} $\mathit{candidates} \geq q-1$
\end{algorithmic}%
}%
\end{algorithm}%
we develop this approach to allow for a total of $\delta$ indels in $S_x[i,j]$
or any ridge in $S_y$, $[1,k]\setminus x$.  In doing so, we make use of two data
structures, $\textsc{Active}^t$ and $\Delta^t$, which are updated and queried
in each iteration for testing interval $i..j$ in $S_x$ with fixed left bound
$i$ and increasing right bound~$j$.  $\textsc{Active}^t[y]$ is a binary vector
that captures all ridges in $S_y$ containing characters from $\mathcal C(S_x[i,j])$.
$\Delta^t[y]$ is a data structure made of $\delta+1$ vectors that act as counters
for indels in ridges of $S_y$---$\Delta^t[y][d']$ records all ridges in
$S_y$ that accumulated $0 < d' \leq \delta$ indels.
Clearly, if $d^*=\textsc{Ridge}_{xy}^c[j]-\textsc{Ridge}_{xy}^c[i]$ trivial indels
have been observed in $S_x[i,j]$ (line~\ref{alg:l_d_indels}), then only
$\delta-d^*$ indels are allowed to occur in any ridge in $S_y$.
Algorithm~\ref{alg:macsi_filter} uses the $(\delta+1-d^*)^{\textit{th}}$ vector
in $\Delta^t[y]$ as a \emph{mask} to ignore all ridges that exceed $\delta$
indels in subsequent calculations.

At each iteration $j$, ridges of $S_y$ can be classified in three categories:
(i) ridges in $\textsc{Ridge}^t_{xy}[j]$ previously observed in any of the
$i..j-1$ iterations; (ii) previously observed ridges not in
$\textsc{Ridge}^t_{xy}[j]$; (iii) new ridges of $\textsc{Ridge}^t_{xy}[j]$.
The last two categories require an update of $\Delta^t$ and hence are assigned
to vectors $a_{\mathit{off}}$ and $a_{\mathit{new}}$, respectively
(lines~\ref{alg:l_aoff}, \ref{alg:l_anew}).  The counter for indels of all ridges
of $a_{\mathit{off}}$ gets increased by one, whereas for $a_{\mathit{new}}$
the counter gets increased by $\min(j-i, \delta)$; both updates occur in
the \textbf{for} loop on lines~\ref{alg:l_delta_update_start}-\ref{alg:l_delta_update_end}. 
After updating $\textsc{Active}^t[y]$ vector
(line~\ref{alg:l_active_update}), Algorithm~\ref{alg:macsi_filter} can test whether
there exists at least one ridge in $S_y$ that satisfies the indel constraint
with $S_x[i,j]$ (lines~\ref{alg:l_indel_test_start}-\ref{alg:l_indel_test_end}).
Algorithm~\ref{alg:macsi_filter} returns \textbf{True} only if $q-1$
strings fulfill the indel constraint; otherwise, the
current position $j$ in reference string $S_x$ gives rise to set $J = [i,j]$
(Algorithm~\ref{alg:macsi}, line \ref{alg:l_det_j}). 

Further improvements in speed are made by individually refining the set of
characters $J$ for each position of set $P_y$, which is constructed in
line~\ref{alg:l_con_p} in Algorithm~\ref{alg:macsi}.  In doing so, we improve on
a technique already described in~\cite{Doerr:2014fq}.
Note that $P_y$ is not
dependent on the candidate set of right bounds $J$ of intervals in $S_x$, but
rather on the left candidate-interval bound $i$ and therefore can be
computed prior to iterating through set~$J$. For each position $p\in P_y$ of
indeterminate string $S_y$, we collect indices $J_p\subseteq J$ in all sets
$\textsc{Pos}_{yx}[k^*]..\textsc{Pos}_{yx}[l^*]$ such that $k^*$ and $l^*$,
$k^*\leq p\leq l^*$, are at most $\delta$ trivial indels from $p$ afar,
i.e.~$J_p = (\textsc{Pos}_{yx}[k^*]\cup\cdots\cup\textsc{Pos}_{yx}[l^*])\cap J$.
Thus we use Algorithm~\ref{alg:macsi_filter} on each position $k'$
of indeterminate string $S_y$ that is visited while walking from $p$ towards
$k^*$. In doing so, we test for the possible existence of AWCI sets that have
members in at least $q-2$ strings other than $S_y$ and $S_x$ and contain
some interval $[k,l]_{S_y}$, $k\leq k'\leq p\leq l$.
We stop
progressing towards $k^*$ if Algorithm~\ref{alg:macsi_filter} returns
\textbf{False}. This process also applies to positions $l'$, $p\leq l'\leq l^*$. 

Just as described in~\cite{Doerr:2014fq}, sets $J_p$, $p\in P_y$, can then be
used to refine the candidate set $J$ of right interval bounds of~$S_x$.
Unlike in~\cite{Doerr:2014fq}, however, the refinement is performed across multiple
sequences. For any set of positions $P_y$, we identify the rightmost interval
bound found in the neighborhood of any $p\in P_y$ that satisfies the indel
treshold in $S_x$, i.e.,
  $$j^\star_y := \max_{p \in P_y}\{j_p~|~|J_p \cap [i,j_p]|\leq \delta\}$$
Then the new candidate set of right interval bounds of the
next iteration is
  $$J = \big[i,\max_{\substack{Z \subseteq [1,m] \setminus x\\|Z| \geq q-1}} \big(\min_{z \in Z} j^\star_z\big)\big]$$
We repeat this process to refine each of the sets $J_p$.
In practice, fewer than 3 iterations suffice to converge to stable sets $J$ and $J_p$. 

\section{Results and Discussion}
\label{sec:results}
We demonstrate the applicability of MACSI in comparing genomic sequences on a
dataset of 93 bacterial genomes that was also used in evaluating the algorithms
for discovering pairs of AWCIs presented in Doerr~\etal~\cite{Doerr:2014fq} and
that were originally obtained from Ciccarelli~\etal~\cite{Ciccarelli:2006}.
Doerr~\etal~transformed the genomic sequences into indeterminate strings, by
using annotions of protein-coding genes and by computing pairwise BLASTP hits
between genes. The resulting indeterminate strings have an average length of
$2726$ elements and each pair of indeterminate strings contains $6499$ positions
with intersecting character sets.

MACSI is implemented in Python; our implementation allows for parallel computation,
but for the purpose of assessing the runtime of our method, all computations were
performed in a single thread.  In addition to indel parameter $\delta$ and quorum,
our implementation allows to set a minimum size for reported AWCIs, where size is
measured between pairs of AWCIs by the number of positions that contain shared
characters. Throughout our experiments, we set the minimum size to $10$.  
\begin{figure}[tb]%
    \includegraphics[width=0.31\columnwidth]{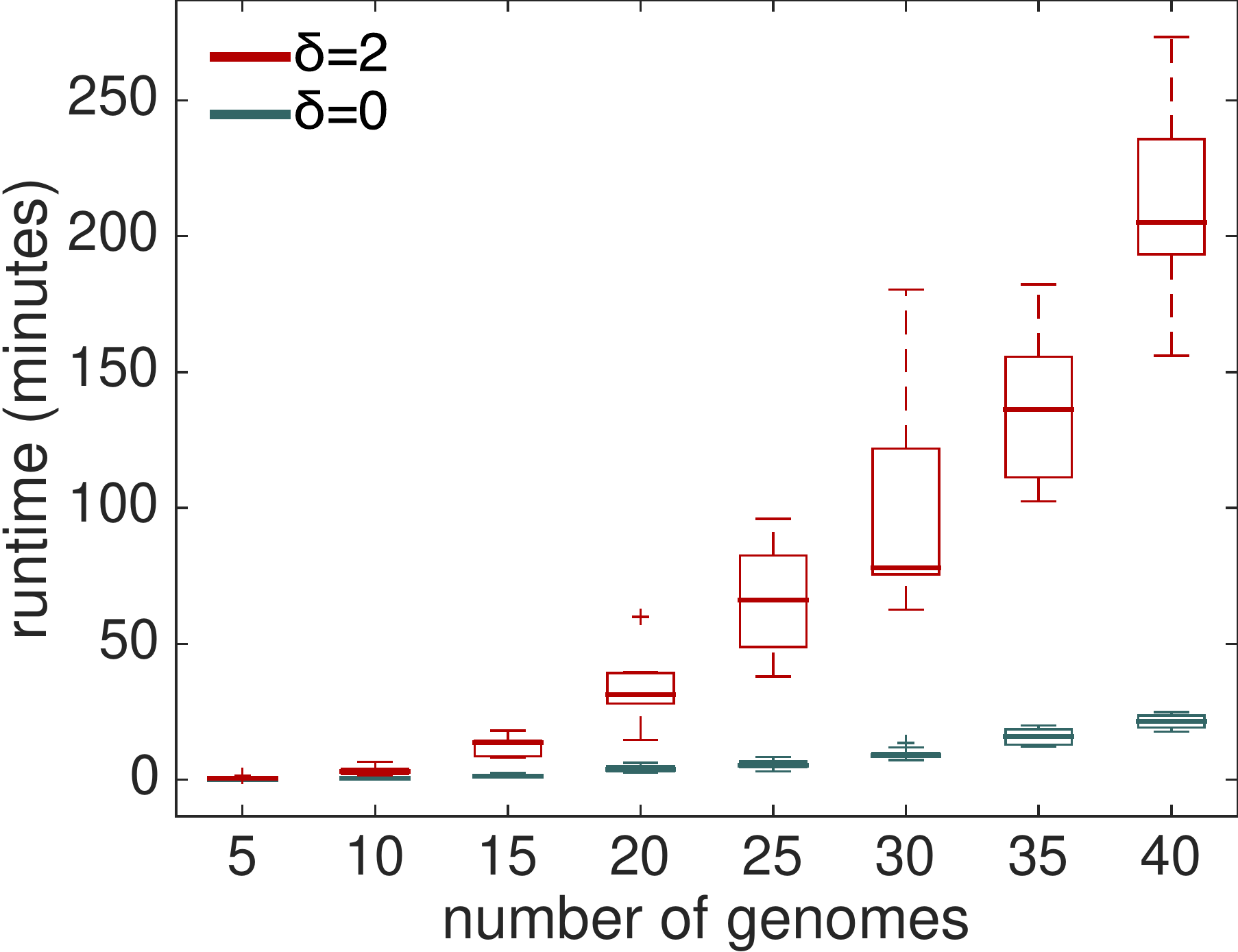}
    \includegraphics[width=0.31\columnwidth]{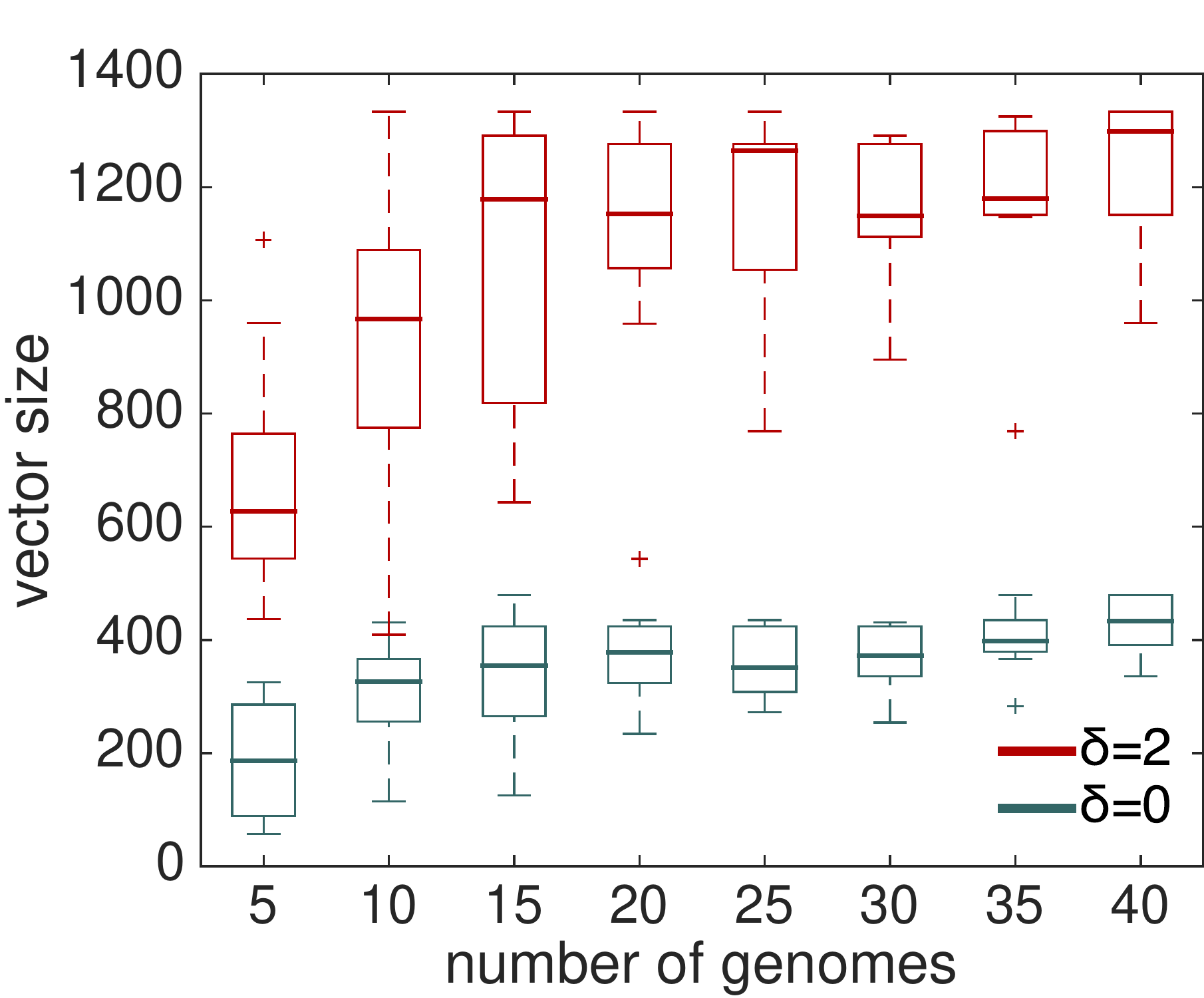}
    \includegraphics[width=0.31\columnwidth]{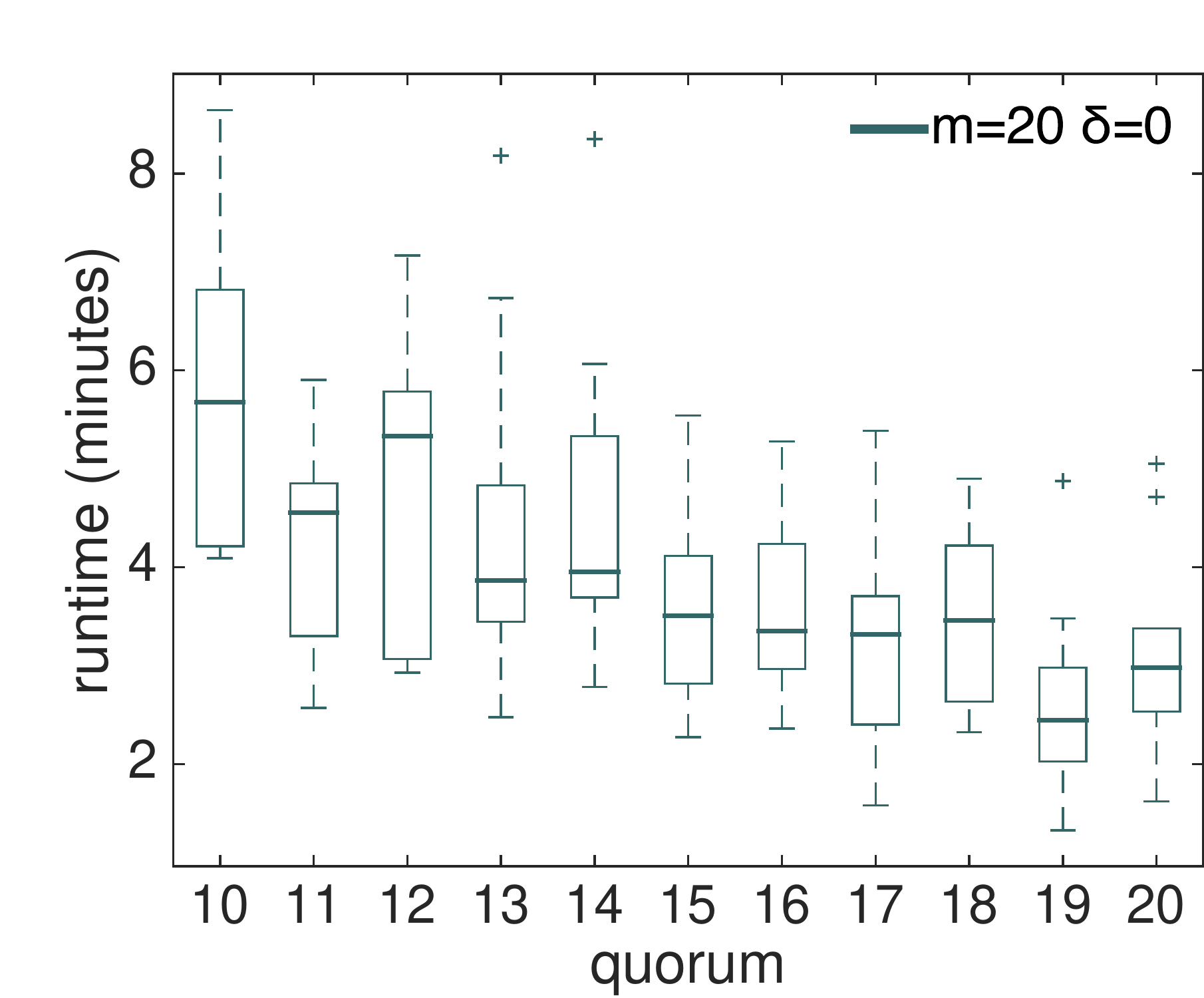}
    {\scriptsize \hfill(a)\hfill\phantom{.}\hfill(b)\hfill\phantom{.}\hfill(c)\hfill\phantom{.}}
    
    \caption{\textbf{(a)} Runtimes of MACSI and \textbf{(b)} bit vector sizes
    (per position) in data structure $\textsc{Ridge}_{xy}^t$ on genome sets of
    different sizes with indel thresholds $\delta \in \{0,2\}$; \textbf{(c)}
    Runtimes of MACSI with indel threshold $\delta=0$ on sampled sets of 20
    genomes with varying quorum settings.}%
    \label{fig:eval}%
\end{figure}%
Experiments were run on a machine with sixty-four 2.3 GHz cores. To assess the
running time of our implementation we sampled genome sets of different sizes
from the pool of 93 bacterial genomes and then ran MACSI using different parameter
settings.  The sampling was repeated 10-fold in all experiments. 
Figure~\ref{fig:eval} shows the results.  Since our speed-up relies on the
frequency of indels, the indel threshold $\delta$ has the most dramatic effect
on the running time of the algorithm; in comparison, lowering the quorum has
only a mild effect.  Comparing 40 full bacterial genomes took on average just around
3.5 hours of computation. 

The number of reported AWCI pairs as well as the data structure
$\textsc{Ridge}_{xy}^t$ have the most decisive effect on the space requirements
of our algorithm.  Because the former is less prone to algorithmic improvements,
we monitored only the latter in our experiments.
Figure~\ref{fig:eval}~(b) illustrates the sizes of bit vectors in
$\textsc{Ridge}_{xy}^t$ in our experiments and shows that fewer than 500 bits
were needed in runs with $\delta=0$ and fewer thatn 1400 bits with $\delta=2$.

\section{Conclusion}
\label{sec:conclusion}
We presented a fast algorithm for the discovery of approximate weak common
interval in multiple indeterminate strings.  Our main contributions are
data structures that allow for efficient filtering of intervals that
cannot be members of AWCI sets of a given minimal size.  We exploited
genomic properties, in particular the fact that our indeterminate
strings will exhibit many trivial indels, yielding easily identifiable
boundaries for AWCIs.  How to speed up the general algorithm in the absence
of trivial indels remains an open question.



\section*{Competing Interests}
  The authors declare that they have no competing interests.

\section*{Author's Contributions}
    BM initiated and directed the research project, DD designed and implemented
    the herein presented algorithms and ran the experiments; both authors
    wrote the manuscript, and read and approved its final version.

%

%
\bibliographystyle{abbrv}
\bibliography{main}

\begin{thebibliography}{1}

\bibitem{Ciccarelli:2006}
F.~D. Ciccarelli, T.~Doerks, C.~von Mering, C.~J. Creevey, B.~Snel, and
  P.~Bork.
\newblock {Toward automatic reconstruction of a highly resolved tree of life.}
\newblock {\em Science}, 311(5765):1283--1287, 2006.

\bibitem{Didier:2007hj}
G.~Didier, T.~Schmidt, J.~Stoye, and D.~Tsur.
\newblock {Character sets of strings}.
\newblock {\em J. Discr. Alg.}, 5(2):330--340, 2007.

\bibitem{Doerr:2014fq}
D.~Doerr, J.~Stoye, S.~B{\"o}cker, and K.~Jahn.
\newblock Identifying gene clusters by discovering common intervals in
  indeterminate strings.
\newblock In {\em Proc.\ 12th RECOMB Satellite Workshop on Comparative Genomics
  RECOMB-CG'14}, volume 15 (Suppl. 6) of {\em BMC Genomics}, page~S2, 2014.

\bibitem{Ghiurcuta:2014dx}
C.~Ghiurcuta and B.~Moret.
\newblock Evaluating synteny for improved comparative studies.
\newblock In {\em Proc.\ 22nd Symp. on Intelligent Systems for Mol.\ Bio.\
  ISMB'14}, volume 30(12) of {\em Bioinformatics}, pages i9--18, 2014.

\bibitem{Uno:2000ca}
T.~Uno and M.~Yagiura.
\newblock {Fast algorithms to enumerate all common intervals of two
  permutations}.
\newblock {\em Algorithmica}, 26(2):290--309, 2000.

\bibitem{Visnovska:2013ua}
M.~Visnovsk{\'a}, T.~Vina{\v{r}}, and B.~Brejov{\'a}.
\newblock Dna sequence segmentation based on local similarity.
\newblock In {\em Proc.\ 13th Conf.\ on Info.\ Technologies---Applications and
  Theory ITAT'13}, pages 36--43, 2013.

\bibitem{ZhangMoret}
X.~Zhang and B.~Moret.
\newblock Refining regulatory networks through phylogenetic transfer of
  information.
\newblock {\em IEEE/ACM Trans.\ on Computational Biology and Bioinformatics},
  9(4):1032--1045, 2012.

\end{thebibliography}

\end{document}